\documentclass[preprint]{aastex}

\shorttitle{Comparison of Global Model with Coronal Mass Ejections}
\shortauthors{Yeates et al.}

\begin{document}

\title{Comparison of a Global Magnetic Evolution Model with Observations of Coronal Mass Ejections}

\author{A. R. Yeates\altaffilmark{1}, G. D. R. Attrill}
\affil{Harvard-Smithsonian Center for Astrophysics, 60 Garden Street, Cambridge, MA 02138}
\altaffiltext{1}{Now at Division of Mathematics, University of Dundee, Dundee, DD1 4HN, UK}
\email{anthony@maths.dundee.ac.uk, gattrill@cfa.harvard.edu}

\and
\author{Dibyendu Nandy}
\affil{Department of Physical Sciences, Indian Institute of Science Education and Research, Kolkata, Mohanpur 741252, West Bengal, India}
\email{dnandi@iiserkol.ac.in}

\and
\author{D. H. Mackay}
\affil{School of Mathematics and Statistics, University of St. Andrews, St. Andrews, KY16 9SS, UK}
\email{duncan@mcs.st-and.ac.uk}

\and
\author{P. C. H. Martens, A. A. van Ballegooijen}
\affil{Harvard-Smithsonian Center for Astrophysics, 60 Garden Street, Cambridge, MA 02138}
\email{pmartens@cfa.harvard.edu,vanballe@cfa.harvard.edu}

\begin{abstract}
The relative importance of different initiation mechanisms for coronal mass ejections (CMEs) on the Sun is uncertain. One possible mechanism is the loss of equilibrium of coronal magnetic flux ropes formed gradually by large-scale surface motions. In this paper, the locations of flux rope ejections in a recently-developed quasi-static global evolution model are compared with observed CME source locations over a 4.5-month period in 1999. Using EUV data, the low-coronal source locations are determined unambiguously for 98 out of 330 CMEs. An alternative method of determining the source locations using recorded H$\alpha$ events was found to be too inaccurate. Despite the incomplete observations, positive correlation (with coefficient up to 0.49) is found between the distributions of observed and simulated ejections, but only when binned into periods of one month or longer. This binning timescale corresponds to the time interval at which magnetogram data are assimilated into the coronal simulations, and the correlation arises primarily from the large-scale surface magnetic field distribution; only a weak dependence is found on the magnetic helicity imparted to the emerging active regions. The simulations are limited in two main ways: they produce fewer ejections, and they do not reproduce the strong clustering of observed CME sources into active regions. Due to this clustering, the horizontal gradient of radial photospheric magnetic field is better correlated with the observed CME source distribution (coefficient 0.67). Our results suggest that, while the gradual formation of magnetic flux ropes over weeks can account for many observed CMEs, especially at higher latitudes, there exists a second class of CMEs (at least half) for which dynamic active region flux emergence on shorter timescales must be the dominant factor. Improving our understanding of CME initiation in future will require both more comprehensive observations of CME source regions and more detailed magnetic field simulations.
\end{abstract}


\section{Introduction}

Coronal mass ejections (CMEs) are a major area of interest in solar physics, both because of their influence on the near-Earth environment \citep{gosling1974,schwenn2006}, and because of their role in the global magnetic field evolution \citep{bieber1995,zhang2001,owens2007}. However, the physical mechanism or mechanisms responsible for their initiation remains an open question; while there is agreement that the driving energy must originate in the magnetic field, the manner in which this free magnetic energy is built up and released is still under debate \citep{forbes2000,klimchuk2001,low2001}.

The present study considers a particular model for CME initiation: the quasi-static build-up and sudden loss of equilibrium of coronal magnetic flux ropes, in response to the large-scale surface motions of differential rotation and meridional flow, and to flux cancellation. These motions generate electric currents in the corona and the resulting magnetic helicity becomes concentrated in twisted flux rope structures above polarity inversion lines in the photospheric field \citep{pneuman1983,vanballegooijen1989}. This formation process takes place over timescales on the order of a month and is therefore an appealing mechanism for the formation of quiescent filaments \citep{vanballegooijen1990,zirker1997,vanballegooijen1998,mackay2005,gibson2006}. Recent simulations of the global corona demonstrate that this mechanism can account for the observed hemispheric pattern of filament chirality \citep{yeates2008a,yeates2009a}, as well as leading to the ejection of flux ropes \citep{yeates2009b}. On shorter timescales (hours to days), observations show that magnetic flux in active regions can emerge highly twisted, carrying substantial currents from beneath the photosphere \citep[e.g.,][]{leka1996,lites2009}. Although we include a net magnetic helicity in emerging active regions, our large-scale, quasi-static model cannot follow the dynamic evolution of the flux emergence and its rapid reconfiguration on entering the corona. Indeed, the model was originally designed to follow the large-scale dispersal of magnetic flux and helicity from active regions, so does not, at present, follow the detailed evolution on short timescales that is observed inside active regions. This paper aims to determine how the distribution of flux rope ejections in the model compares with that of observed CMEs. We do this through a direct comparison between the simulation results of \citet{yeates2009b} and CME observations over a 4.5-month period in 1999 using the Solar \& Heliospheric Observatory (SOHO). This is the first such comparison of CMEs with long-term simulations representing observed magnetic configurations on the Sun.

A difficulty with such a direct observational comparison lies in the identification of CME source regions in the low corona. There are multiple low-coronal signatures that may indicate the CME source region, including: coronal dimming regions, a post-eruptive arcade, a filament eruption and/or a coronal wave.  However, on occasion, CMEs are detected which appear to have no low-coronal source region at all \citep[e.g.][]{robbrecht2009b}.  Such CMEs may originate from higher in the corona, and lack a low-coronal signature.  Even when a low-coronal signature is observed, confidently linking it with a particular CME requires a case-by-case study, and consideration of both spatial location and a plausible temporal association between the CME and its low coronal signature is required. Part of our aim in this paper is therefore to consider how meaningful a comparison of simulations with observations can be, and what would be needed for a better comparison.

The key feature of this work is that we simulate the global magnetic field in the solar corona, so that flux ropes form in a time-dependent manner at different locations on the Sun, in response to emergence of active regions and large-scale surface motions. This is in contrast to previous studies which have typically modelled a single CME event in a simplified magnetic configuration, in order to consider the basic physical processes leading to loss of equilibrium. These studies indicate that the evolution of flux ropes, and whether sudden eruption will occur, depends both on the photospheric footpoint motions \citep{forbes1991,mikic1994,amari1996,amari2003,mackay2006a} and on the overlying, background magnetic field \citep{isenberg1993,antiochos1999,lin2005}. In the model used in this paper, the footpoint motions are determined by the large-scale surface motions of differential rotation, meridional flow, and supergranular diffusion, while the overlying magnetic field at the location of each flux rope is determined self-consistently in the global magnetic configuration. This allows us to place constraints on the applicability of this CME initiation mechanism on the real Sun. In this context, several recent studies have looked at the large-scale magnetic topology of observed CME source regions, focusing on whether the background field is bipolar or quadrupolar, because CME initiation models differ fundamentally in this respect \citep{li2006,barnes2007,ugarteurra2007,cook2009}. Our model takes into account this topology automatically, because both the flux rope and its overlying field are part of the global magnetic configuration.

The lack of any previous such global models may be explained by the need for coronal electric currents, required in order to store free magnetic energy \citep[e.g.,][]{schrijver2008}. These must be built up in a time-dependent manner, either rapidly through the emergence of pre-twisted structures, or over longer timescales by surface shearing, as in the simulations in this paper. Global full-MHD models are too computationally demanding to model this time evolution over many weeks, although they have been successful in modelling both global equilibria \citep[][]{riley2006,cohen2007,lionello2009} and the evolution of individual CME events \citep[][]{riley2008,manchester2008,cohen2009}. Instead, our model uses a quasi-static approximation to the magnetic field evolution in order to follow the formation of currents and the transport of magnetic helicity, albeit in a simplified manner.

In Section \ref{sec:obs} we describe the instrumentation and data reduction used to identify the source regions of observed CMEs. The main features of the simulations are outlined in Section \ref{sec:sims}, before comparing the observed and simulated distributions of CME sources in Section \ref{sec:comp}. The relation between the two is discussed in Section \ref{sec:discuss}, and conclusions in Section \ref{sec:conclusion}, where we also make recommendations for future study.

\section{Observations of CME Source Regions} \label{sec:obs}

We began by compiling a list of CMEs between 1999 May 13 and 1999 September 26 from the CDAW (coordinated data analysis workshops) catalog\footnote{Available online from the CDAW data centre \url{http://cdaw.gsfc.nasa.gov/CME\_list/}.} \citep{yashiro2004}. This is the standard manually-compiled list of CMEs observed by the Large Angle and Spectrometric Coronagraph (LASCO) experiment since 1996 \citep{brueckner1995}. The C2 and C3 coronagraphs observe the white-light corona from  $2 R_\odot$ to $7 R_\odot$ and $3.7 R_\odot$ to $32 R_\odot$ respectively. Our comparison period is selected from the rising phase of the Solar Cycle, so as to include magnetic field structures representative of both Solar Maximum (in the newly-emerged active regions) and Solar Minimum (in the remnant regions and polar fields at higher latitudes). This period has the further advantage that it was used in our previous simulations to investigate the parameter dependence of the model \citep[most recently with regard to the formation and ejection of flux ropes,][]{yeates2009b}. The start date of our comparison period is chosen to allow sufficient time for the simulation to evolve away from the initial condition on 1999 April 16 \citep[see][]{yeates2009b}, while the end date 
just precedes the break in SOHO/EIT observations that occurred between 1999 September 27 and 1999 October 5.

In order to concentrate on well-observed CMEs, we ignored all events labelled ``poor'' by the CDAW observers. After studying the LASCO/C2 running difference movies of each remaining event, a further 12 very weak events were removed, as were 19 events which we could not clearly identify to be independent from other events. In addition, our study of corresponding EIT observations (to be described below) led us to split one event in the CDAW catalog---on 1999 August 18 at 05:54---into two separate eruptions. A list of 330 CMEs remained, including both halo and limb events.

There are some well-known uncertainties in the CDAW observations. Firstly, they are sensitive to projection effects, with CMEs in the plane of the sky being better observed \citep{hudson2006}. However, since our data cover several solar rotations, there should be no systematic bias in the overall longitude distribution. Secondly, the selection of events is subjective. Recently, an automated CME catalog, CACTus \citep{robbrecht2004,robbrecht2009a}, has been developed to reduce subjectivity in the detection of CMEs by detecting radial motion in height-time maps of LASCO data. Although many additional events are detected in CACTus, they tend to be narrow events, background outflows, or multiple detections of the same CME \citep{yashiro2008}. Thus the manual CDAW catalog is sufficient for comparison with simulated flux rope ejections. The third main limitation of LASCO observations is that they do not show the initiation locations of eruptions in the low corona, because CMEs often move non-radially before they reach the C2 field of view \citep[{\it e.g.},][]{plunkett2001, attrill2009}. In this study we have used additional observations in extreme ultraviolet (EUV) to determine, where possible, the CME source locations in the low corona. Our unsuccessful attempt to use an alternative H$\alpha$ data set is described in Section \ref{sec:halpha}.

\subsection{EUV Images} \label{sec:eit}

\begin{figure}[htb]
\includegraphics[width=\textwidth]{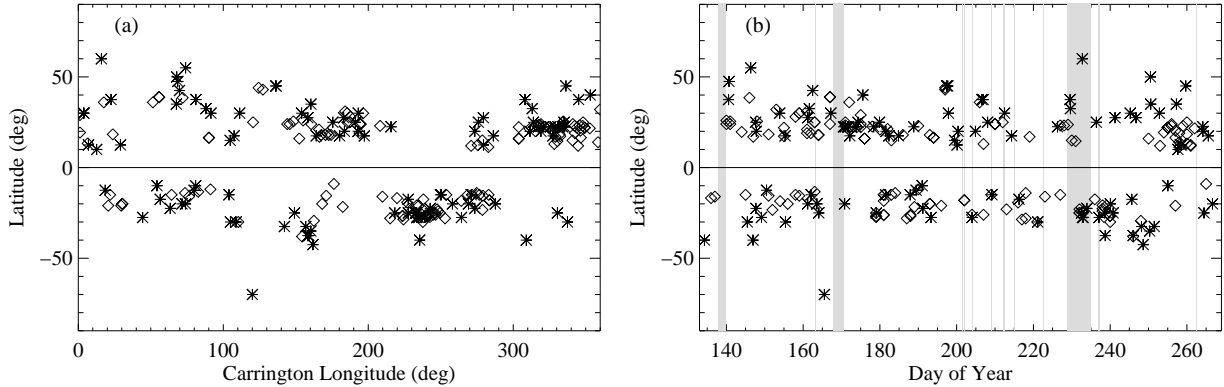}
\caption{Source locations associated with observed CMEs, in (a) latitude-longitude and (b) latitude-time. Asterisks show locations determined from (category 1) EUV observations, and diamonds show locations determined from H$\alpha$ data (Section \ref{sec:halpha}). In (b) light-gray shading denotes LASCO data gaps recorded on the CDAW website.}\label{fig:obs}
\end{figure}

\begin{deluxetable}{lr}
\tablewidth{0pt}
\tablecaption{Numbers of CMEs with associated EUV source locations.}
\tablehead{
\colhead{EIT source category}  & \colhead{Number of CMEs}
}
\startdata
1 (clear) & 98\\
2 (plausible) & 44\\
3 (far-side) & 55\\
4 (no associated EIT source) & 124\\
No EIT data & 9\\
\tableline
Total & 330\\
\enddata
\label{tab:obs}
\end{deluxetable}

A number of low-coronal features associated with CMEs appear in the EUV images from SOHO/EIT \citep[Extreme ultraviolet Imaging Telescope;][]{delaboudiniere1995}. These include erupting filaments, flares, post-eruption arcades, and transient coronal dimmings, in addition to EIT coronal waves \citep{plunkett2001}. EIT observations at up to 12-minute cadence were available in the $195\textrm{\AA}$ filter for the hours leading up to and following 321 of the 330 CMEs in our study period. Each of these were studied manually, in conjunction with the LASCO movie showing the CME. Possible CME signatures were recorded. We were unable to use EIT images in the $171\textrm{\AA}$, $284\textrm{\AA}$, or $304\textrm{\AA}$ filters as images were available only on one or two days during each month. To minimize subjective bias, two of us (ARY and GDRA) carried out this analysis independently, before comparing results and compiling a final event list (given in Appendix \ref{sec:app}). Further, each CME was assigned to one of the following categories:
\begin{enumerate}
\item Clearly associated front-side source visible in EIT.
\item Plausibly associated front-side source visible in EIT.
\item Source becomes visible in EIT above the limb, but originated behind limb.
\item No plausibly-associated EIT source evident. 
\end{enumerate}
The first two categories give two degrees of certainty to our front-side source identifications. The third category describes CMEs where the low coronal source of the CME becomes visible in EIT above the limb, but its source on the solar disk lies behind the limb, so that a longitude position can not be determined. The locations of the possible sources in categories 1 and 2 were recorded manually by overlaying a latitude-longitude grid on the EIT movies. Identified source locations in category 1 are shown by asterisks in Figure \ref{fig:obs}. Table \ref{tab:obs} lists the number of CMEs in each category. For the comparison with simulations in this paper, we use only the 98 source locations in category 1, {\it i.e.}, those clearly identified with CMEs. This represents only 30\% of the observed CMEs, highlighting the difficulties associated with identifying CME source regions in the low corona.

\begin{figure}[!htb]
\includegraphics[width=\textwidth]{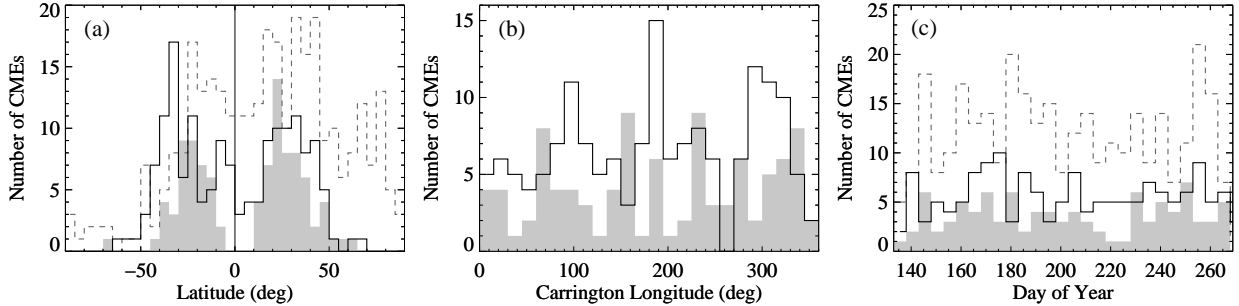}
\caption{Histograms showing (a) latitude, (b) longitude, and (c) time distributions of observed CME sources (gray shaded) and simulated ejections in run A4 (solid lines). Dashed lines show distributions of apparent latitudes and times in the original LASCO/CDAW data. Bin sizes are $5^\circ$ in (a), $15^\circ$ in (b), and $5$ days in (c). }\label{fig:obshist}
\end{figure}

Latitude, longitude and time distributions of CME source regions identified in EUV are shown by the filled gray histograms in Figure \ref{fig:obshist}. The bimodal latitude distribution of EUV sources is consistent with those found over the same period by \citet[][see their Figure 10b]{cremades2004}, who carried out a similar SOHO/EIT analysis for a subset of ``structured'' CMEs between 1996 and 2002. Although our analysis in this paper is restricted to the year 1999, the results of \citet{cremades2004} indicate that the latitude distribution changes over the solar cycle, with an extension of the CME source distribution right down to the equator at solar maximum, in addition to a high latitude branch corresponding to polar crown filament eruptions. This cycle variation is supported by observations of the latitudes of disappearing filaments \citep{pojoga2003}.

The longitude distribution of CME sources is less structured, although from Figure \ref{fig:obs}(a) we see a strong clustering into several activity complexes \citep{gaizauskas1983}, such as at longitude $250^\circ$ in the southern hemisphere, or $330^\circ$ in the northern hemisphere. These active longitudes are also responsible for major fluctuations in the time distribution (Figure \ref{fig:obshist}c).

The dashed histograms in Figures \ref{fig:obshist}(a) and (c) show the apparent latitude and time distributions of all 330 CMEs from the CDAW catalogue. As expected, the distribution differs from those of the low coronal sources \citep{gopalswamy2003}. The two main differences are (1) more events at the equator, consistent with the deflection of some events towards the equator as they propagate out \citep{plunkett2001,attrill2009}, and (2) more events at high latitudes in the Northern hemisphere. This north-south asymmetry in the latitude distribution of LASCO CMEs is a feature of the year 1999, evident in Figure 3 of \citet{yashiro2004}.

The solid black histograms in Figure \ref{fig:obshist} show the equivalent distributions of flux rope ejections in simulation run A4. The simulations will be described in Section \ref{sec:sims}.

\subsection{H$\alpha$ Event Database} \label{sec:halpha}

As an alternative to using EUV images, we also tried identifying source locations of observed CMEs using a database of H$\alpha$ events assembled by \citet{howard2008}. This database lists the locations of flares and disappearing filaments in Solar Geophysical Data that may be associated with CMEs in the CDAW catalogue. A surface event is associated with a particular CME if it occurred within $\pm1$ hour of the CME onset time (estimated assuming a constant outward speed), and if its latitude and longitude are in the same quadrant as any part of the angular span of the CME. It should be noted that, in this database, some CMEs are associated with multiple possible surface events, and many with none. Also, some surface events are listed with multiple CMEs. In an attempt to select only those CMEs with reasonably consistent locations given by the H$\alpha$ events, we filter out CMEs whose list of associated H$\alpha$ events has standard deviations greater than $10^\circ$ in latitude or $20^\circ$ in longitude. This leaves 137 CMEs with consistent locations in the H$\alpha$ database, in the sense that the H$\alpha$ events associated with each CME are reasonably close together spatially. These source locations are shown by diamonds in Figure \ref{fig:obs}. It is clear from Figure \ref{fig:obs} that, like the locations identified in EUV, those identified in H$\alpha$ data are not uniformly distributed, but rather are clustered into several major activity complexes.

Unfortunately, we find that the H$\alpha$ source locations are not consistent with the EUV source locations. Firstly, it is evident from Figure \ref{fig:obs} that the H$\alpha$ sources are even more strongly clustered into active regions, with no sources whatsoever observed at latitudes above about $40^\circ$. The correlation coefficient between the EUV and H$\alpha$ spatial distributions is only $0.68$ (when binned in $20^\circ$ latitude bins and $30^\circ$ longitude bins, following the method described in Section \ref{sec:comp}). The discrepancy becomes even more serious when considering the source locations identified with particular CMEs. There are 42 CMEs with both a category 1 source location in EUV and a location identified in H$\alpha$. However, the H$\alpha$ location overlaps the EUV location (to within $\pm10^\circ$ in latitude and $\pm20^\circ$ in longitude) in only 22 of these cases. We are inclined to favor the EUV locations rather than the H$\alpha$ locations (at least for category 1 EUV events) because the EUV data were examined in detail on a case-by-case basis, which differs from the approach used to compile the H$\alpha$ listing. The less careful association of H$\alpha$ events with individual CMEs is exemplified by 23 events which were clearly seen to originate from behind the limb in EUV, yet in the H$\alpha$ database were associated with flares that happened to occur co-temporally on the solar disk. The H$\alpha$ observations do have the potential advantage of a higher cadence than the EUV images (minutes rather than over $\sim10$ minutes), so could in principle identify additional CME sources that were missed in EUV. However, in this particular database, we have no clear way of selecting which of many possible H$\alpha$ events are actually associated with each (or any) CME. Therefore, we base the comparison with our simulations on the EUV data only.

\section{Coronal Magnetic Field Simulations} \label{sec:sims}

\begin{deluxetable}{crrrrr}
\tabletypesize{\footnotesize}
\tablewidth{0pt}
\tablecaption{Summary of different simulation runs.}
\tablehead{
\colhead{Run} & \colhead{$\beta$ in N. hemisphere} & \colhead{$\beta$ in S. hemisphere} & \colhead{$\eta_0$ ($\textrm{km}^2\,\textrm{s}^{-1}$)} & \colhead{$v_0$ ($\textrm{km}\,\textrm{s}^{-1}$)\tablenotemark{a}} & \colhead{Flux rope ejections per day\tablenotemark{b}}}
\startdata
AN & \multicolumn{2}{c}{No emerging regions} & 45 & 100 & $0.17\pm 0.03$\\
Am6 & 0.6 & -0.6 &  45 & 100 & $1.09\pm 0.16$\\
Am4 & 0.4 & -0.4 &  45 & 100 & $1.02\pm 0.15$\\
Am2 & 0.2 & -0.2 &  45 & 100 & $0.64\pm 0.10$\\
A0 & 0 & 0 &  45 & 100 & $0.72\pm 0.11$\\
A2 & -0.2 & 0.2 &  45 & 100 & $0.99\pm 0.15$\\
A4 & -0.4 & 0.4 &  45 & 100 & $1.15\pm 0.17$\\
A6 & -0.6 & 0.6 &  45 & 100 & $1.27\pm 0.19$\\
D4 & -0.4 & 0.4 & 22.5 & 100 & $1.46\pm 0.22$\\
V4 & -0.4 & 0.4 & 45 & 50 & $1.12\pm 0.17$\\
\enddata
\tablenotetext{a}{Peak value of radial outflow velocity, at $r=2.5R_\odot$.}
\tablenotetext{b}{Number of ejections per day between 1999 May 13 (day of year 133) and 1999 September 26 (day of year 269). Errors are those for the automated flux rope detection \citep[see][]{yeates2009b}.}
\label{tab:runs}
\end{deluxetable}

Our numerical simulations of the large-scale coronal magnetic field evolution were described in detail in \citet{yeates2009b}, and are based on the mean-field model of \citet{vanballegooijen2000}. Briefly, the large-scale mean magnetic field $\mathbf{B}_0=\nabla\times\mathbf{A}_0$ evolves via the non-ideal induction equation
\begin{equation}
\frac{\partial\mathbf{A_0}}{\partial t} = \mathbf{v}_0\times\mathbf{B}_0 - \eta\mathbf{j}_0,
\end{equation}
where the turbulent diffusivity $\eta$ is given by a background value $\eta_0$ and an enhancement in regions of strong current density \citep{mackay2006a}. The velocity is determined by the magneto-frictional technique \citep{yang1986} as
\begin{equation}
\mathbf{v}_0 = \frac{1}{\nu}\frac{\mathbf{j}_0\times\mathbf{B}_0}{B^2} + v_\textrm{out}(r)\hat{\mathbf{r}},
\label{eqn:v0}
\end{equation}
where the first term approximates the evolution as a sequence of quasi-static force-free equilibria, in response to flux emergence and shearing by large-scale surface motions on the lower, photospheric, boundary. The surface motions are given by the standard surface flux transport model \citep{sheeley2005}. The second term in equation (\ref{eqn:v0}) represents a radial outflow imposed only near the upper boundary ($r=2.5R_\odot$), to represent the effect of the solar wind in radially opening magnetic field lines.

In this study we use the simulation runs described in \citet{yeates2009b}. All cover the same time period, starting on 1999 April 16. The initial condition is a potential-field source-surface extrapolation taken from a synoptic normal-component magnetogram from the National Solar Observatory, Kitt Peak, corrected for differential rotation \citep{yeates2007}. The surface and coronal magnetic fields are then evolved continuously for 177-days. During this evolution, 119 new bipolar magnetic regions are inserted into the simulation, with location, size, tilt angle and magnetic flux determined from Kitt Peak synoptic magnetograms \citep{yeates2007}. The bipolar regions take the idealized mathematical form given in \citet{yeates2008a}. Because the simulation is non-potential, non-zero currents and magnetic helicity are generated in the corona during the evolution. This arises not only due to shearing by the large-scale surface motions, but also because the newly-emerging bipolar regions may be given a non-zero helicity, controlled by a ``twist'' parameter $\beta$, described in detail by \citet{yeates2009b}. As discussed in that paper, although techniques to measure the twist in observed active regions have been developed \citep[see][]{nandy2008}, available observations are limited such that we cannot reliably determine the amount of helicity and thus optimum value of $\beta$ to model each individual active region. For simplicity, we have assumed that all bipoles in each hemisphere have the same value of $\beta$, but with different values of $\beta$ in each hemisphere \citep[taking into account the hemispheric dependence of helicity observed by][]{pevtsov1995}. We have run a series of simulations with different values of $\beta$ to study the effect of this parameter on the formation and evolution of magnetic flux ropes. The different runs are summarized in Table \ref{tab:runs}, repeated for convenience from \citet{yeates2009b}.

A natural consequence of the quasi-static evolution simulated in this model is the accumulation above polarity inversion lines of axial magnetic field, in the form of twisted magnetic flux ropes \citep{vanballegooijen1989}. The formation of flux ropes in the mean-field model has been studied in detail for a simple 2-bipole configuration \citep{mackay2006a}, and more recently in the global simulations used in this paper \citep{yeates2009b}. In these simulations it is found that, once the axial magnetic field in a flux rope grows too strong relative to the overlying field, the flux rope loses equilibrium, rises, and is ejected through the top boundary of the computational domain. It is these flux rope ``ejections'' that we compare in this paper to observed low-coronal CME source regions.

\begin{figure*}[!htb]
\includegraphics[width=\textwidth]{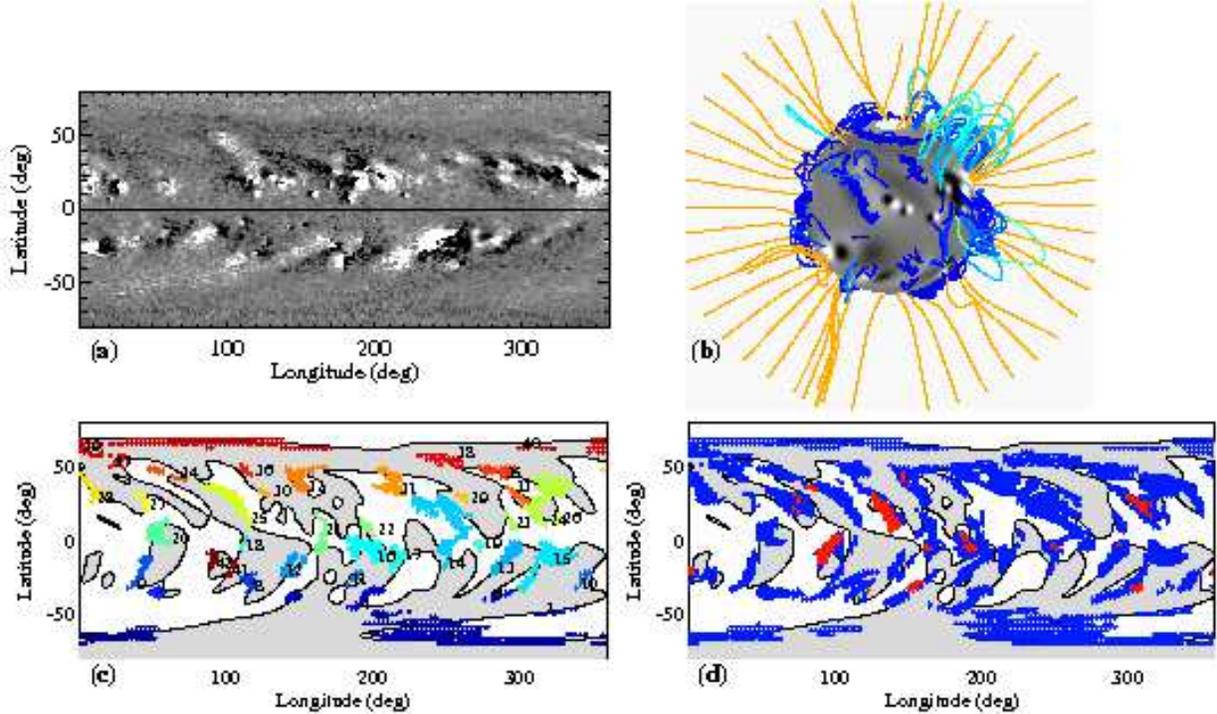}
\caption{Overview of the simulated magnetic field and flux rope detection: (a) normal-component magnetogram from NSO/Kitt Peak for CR1953; (b) snapshot of the simulated magnetic field on day 249; (c) locations of magnetic flux ropes detected on day 240; (d) locations of flux rope points (blue) and ejections (red) detected between days 235 and 245.}\label{fig:sim}
\end{figure*}

\citet{yeates2009b} developed automated methods to identify flux rope structures and their ejection in the global simulations, allowing objective comparison between different simulation runs. In Figure \ref{fig:sim} we show the results of this procedure for one particular day (1999 August 28, day of year 240), in simulation run A4. Figure \ref{fig:sim}(b) shows the simulated magnetic field viewed from Carrington longitude $130^\circ$, with the radial magnetic field on the solar surface shown in grayscale and coronal field lines in blue (if closed) or orange (if open). For comparison, the Kitt Peak magnetogram for CR1953 is shown in Figure \ref{fig:sim}(a). Most field lines in Figure \ref{fig:sim}(b) have been traced from the flux rope points selected by the automated technique \citep[see][]{yeates2009b}. These flux rope points are shown projected on the solar surface in Figure \ref{fig:sim}(c), where each flux rope structure is numbered. The background shading again shows the radial surface magnetic field on the same day (here white is positive and gray negative). Figure \ref{fig:sim}(d) superimposes all flux rope points selected between day 235 and day 245, with the points colored red if they are involved in an ejection during this period, blue otherwise.

The number of ejections in the simulations is found to depend (to some extent) on the twist of emerging bipolar regions, with more ejections for $\beta$ of greater magnitude, or with the observed majority sign in each hemisphere. The right-most column of Table \ref{tab:runs} shows the rate of flux rope ejections in each simulation run, over the period selected for the observational comparison in this paper (1999 May 13 to 1999 September 26). In the next section, we compare the distribution of these flux rope ejections with observed CME source locations.

\section{Comparison between Simulations and Observations} \label{sec:comp}

\begin{figure*}[!htb]
\includegraphics[width=\textwidth]{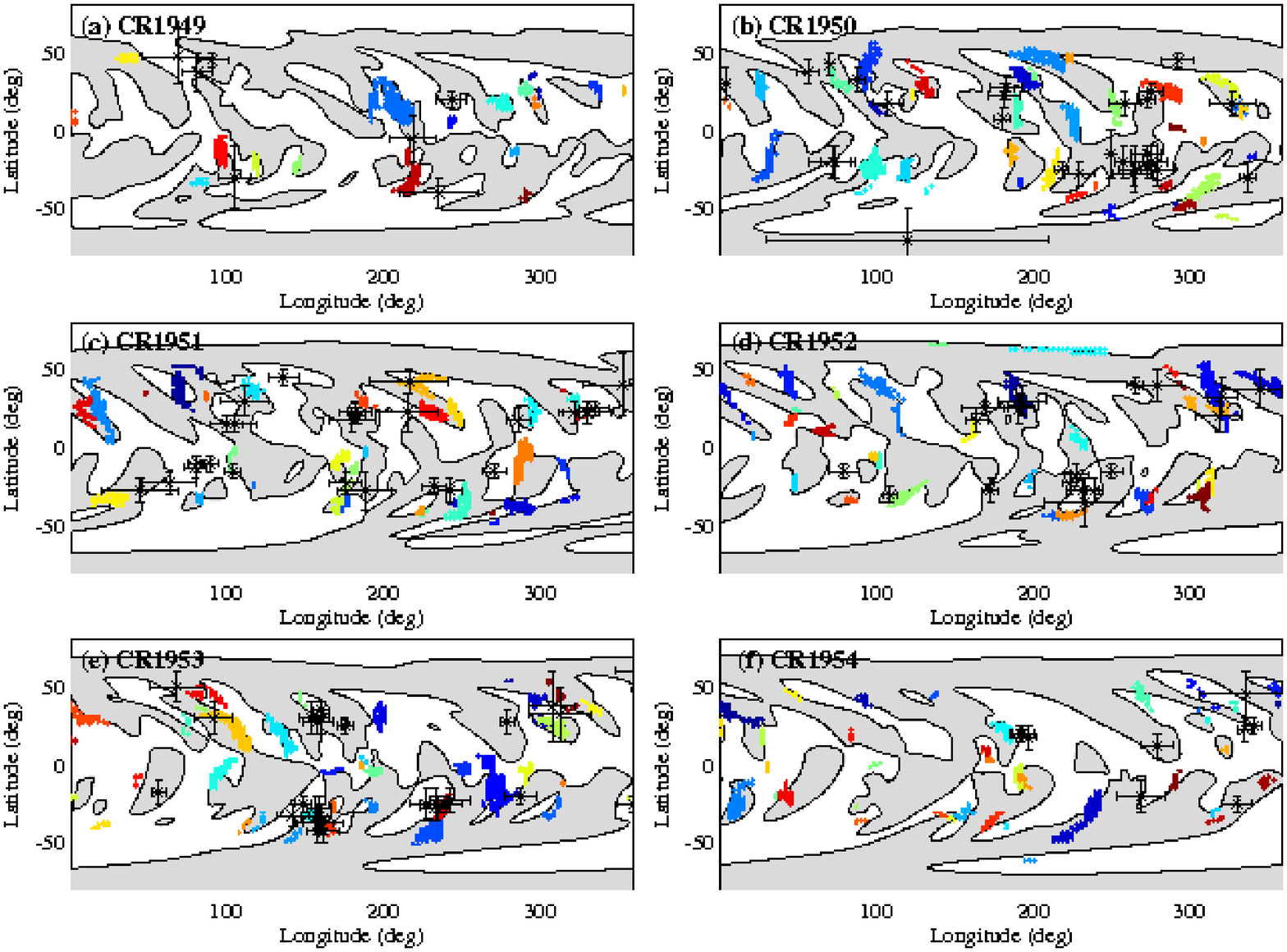}
\caption{Carrington maps showing locations of simulated ejections in run A4 (clusters of colored points), and category 1 CME source locations observed in EUV (with crosses showing the approximate extent of the source region). The polarity of the photospheric radial magnetic field in the simulation is shown in the background (white for positive, gray for negative), for the middle day of each Carrington rotation.}\label{fig:maps}
\end{figure*}

Having identified the locations both of a subset of observed CME sources (Section \ref{sec:obs}) and of simulated flux rope ejections (Section \ref{sec:sims}), we now compare the distributions of each in latitude, longitude and time. Of course, given the limited number of observed CMEs where sources could be reliably determined, this comparison is necessarily approximate.

Figure \ref{fig:maps} shows the locations of observed CMEs and simulated ejections (run A4), for each Carrington rotation during the simulation period. Here the crosses show locations and extents of EIT category 1 CME source locations, while clusters of different colored points refer to different flux rope ejections in the simulation. Each color is equivalent to a separate cluster of red points in Figure \ref{fig:sim}(d). Note that we only compare simulated ejections falling within the comparison period, which begins midway through CR1949 and ends midway through CR1954, hence the smaller number of observed CMEs shown in Figures \ref{fig:maps}(a) and (f). Figure \ref{fig:maps} shows that the simulated ejections clearly do not match the observed locations on a one-to-one basis. This is not surprising, given both the limited observations and poorly constrained simulation parameters such as the twist of emerging bipolar regions, their date of emergence, or the turbulent diffusivity in the corona. However, the locations of many simulated and observed ejections do overlap, and there is a region in the Southern hemisphere around Carrington longitudes $150^\circ$ with few ejections in either observations or simulations. Furthermore, the overall latitude distribution for run A4 is bimodal and broadly consistent to  that of the observed source locations (Figure \ref{fig:obshist}a).

To quantify the association between observed and simulated distributions, we carry out a straightforward correlation analysis. Each list of ejection locations (simulated or observed) is binned in latitude, longitude and time. The bin sizes in latitude and longitude are chosen so as to take into account the spatially-extended nature of the source regions, which are not single points. From cumulative distributions of the latitudinal and longitudinal extents of simulated ejections and observed source regions, we select bin sizes of $20^\circ$ in latitude and $30^\circ$ in longitude, so as to be larger than 80\% of events. The variation of bin size in time will be considered below.

\begin{figure}[!htb]
\includegraphics[width=0.5\textwidth]{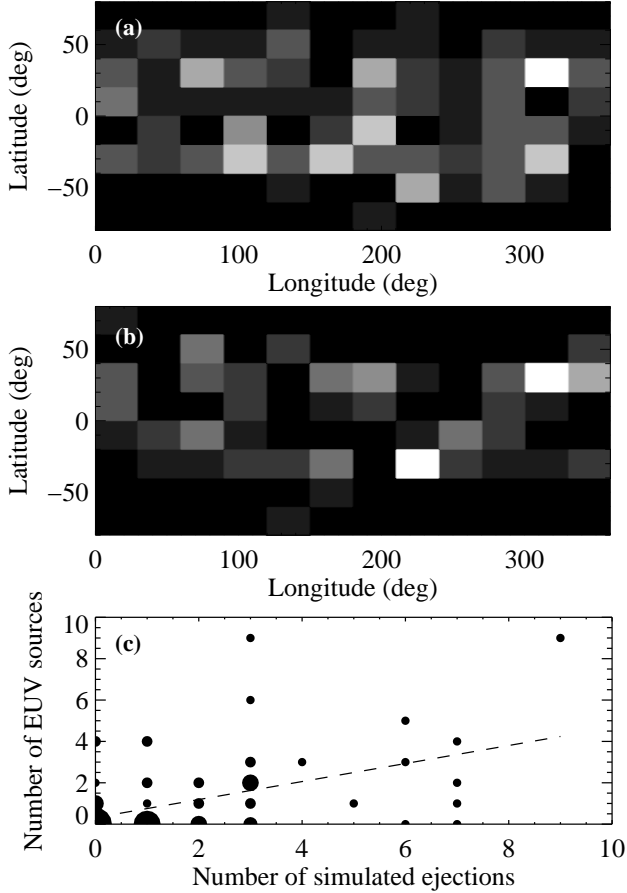}
\caption{Example analysis of correlation in spatial distributions. Histogram in (a) shows the distribution of simulated ejections (run A4), and (b) shows the distribution of observed EUV sources. The correlation between the number of events in each bin is shown in (c), where the sizes of the circles are proportional to the number of bins containing those numbers. The linear correlation coefficient is 0.49.}\label{fig:cormethod}
\end{figure}

To illustrate the technique of estimating a quantitative correlation, the binned distribution of source locations for simulation run A4, in latitude and longitude, is shown in Figure \ref{fig:cormethod}(a). Here black indicates no events in that bin (over the whole 136 days), with white indicating the most events. Figure \ref{fig:cormethod}(b) shows the equivalent distribution for the observed CME sources. To assign a quantitative correlation, we compare the number of simulated ejections with the number of EUV sources in each bin. This is shown by the scatterplot in Figure \ref{fig:cormethod}(c), where the size of each circle indicates the number of bins with those numbers of simulated and observed sources. The Pearson linear correlation coefficient is then computed. In this case it is 0.49, indicating a significant, but not particularly strong, positive correlation.

\subsection{Spatial distributions}

\begin{deluxetable}{cr}
\tablewidth{0pt}
\tablecaption{Correlation between spatial distributions of observed CME sources and simulated ejections.}
\tablehead{
\colhead{Run} & \colhead{Correlation coefficient} }
\startdata
AN  &  0.08 \\
Am6 & 0.37\\
Am4 & 0.33\\
Am2 & 0.33\\
A0 & 0.33\\
A2 & 0.42\\
A4 & 0.49\\
A6 & 0.43\\
D4 & 0.31\\
V4 & 0.40\\
\enddata
\label{tab:spatial}
\end{deluxetable}

Table \ref{tab:spatial} shows the linear correlation coefficients between the binned latitude-longitude distributions of flux rope ejections in the various simulation runs and that of the observed CME sources. All simulation runs show a positive correlation except for run AN (with no emerging bipoles). The correlations are stronger for the simulation runs with the observed majority sign of emerging bipole twist in each hemisphere (negative in the Northern hemisphere and positive in the South, {\it i.e.}, runs A2, A4, A6 and V4) than for the runs with either untwisted bipoles (A0) or bipoles with the opposite sign of twist (runs Am2, Am4 and Am6). However, this is a secondary effect and the positive correlation is present in all runs except AN (where there are no emerging bipoles). The highest correlation coefficient attained is 0.49 for run A4. Interestingly, this is the run that best agreed with observations of filament chirality in our earlier comparison \citep{yeates2008a}. Although a correlation of 0.49 may seem low, it is still significant, given the large-scale simplified nature of the simulations and the uncertainties associated with the observations.

\subsection{Correlation in space and time} \label{sec:time}

\begin{figure}[!htb]
\includegraphics[width=0.5\textwidth]{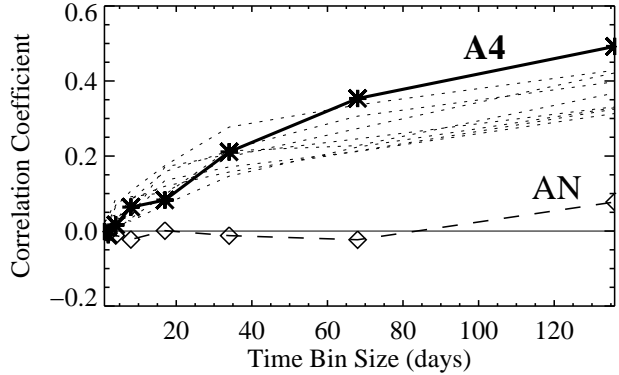}
\caption{Linear correlation coefficients between simulated ejection locations and observed CME source locations as a function of bin size in time. Asterisks/solid line show simulation run A4, diamonds/dashed line show run AN, and dotted lines show other simulation runs.}\label{fig:time}
\end{figure}

Given that the overall latitude-longitude distributions of simulated flux rope ejections and observed CME sources are positively correlated (though not very strongly), are these correlations maintained if the data are also binned in time? Figure \ref{fig:time} shows how the correlation coefficients decrease as the bin size in time is reduced from the whole comparison period (136 days)---which corresponds to the spatial correlations in Table \ref{tab:spatial}---to shorter and shorter bins. We see that significant positive correlations between the simulated ejections and the observed sources are found only with bin sizes of 34 days or longer. This is readily understandable because the observational input driving the simulations---synoptic magnetogram data---is only updated every 27 days. Clearly, if the model is to reproduce observations on a shorter timescale then more frequent updating of emerging flux would be required.

Note that the timescale of flux rope formation and loss of equilibrium in our model also depends on the turbulent diffusivity in the corona, which is not directly constrained by observations. In run D4 we halve the value of the coronal diffusivity, which results in larger flux ropes and 25\% more ejections. However, as pointed out in \citet{yeates2009b}, many of the flux ropes in run D4 are rather highly twisted to be realistic. Furthermore, Table \ref{tab:spatial} shows that the spatial distribution of simulated ejections in run D4 is less well correlated with observed CME source locations (with a correlation coefficient of 0.31, compared with 0.49 for run A4).

\section{Discussion} \label{sec:discuss}

\begin{deluxetable}{rrr}
\tablewidth{0pt}
\tablecaption{Correlation between observed CME sources and simulated surface magnetic field.}
\tablehead{
\colhead{Time bin size (days)} & \colhead{Correlation with $\langle \left|B_r\right|\rangle$} &  \colhead{Correlation with $\langle \left|\nabla_h B_r\right|\rangle$}}
\startdata
136 (full period) & 0.62 & 0.67\\
68 & 0.50 &  0.53\\
34 & 0.41 &  0.43\\
17 & 0.30 &  0.33\\
8 & 0.23 &  0.24\\
4 & 0.18 & 0.19\\
\enddata
\label{tab:br}
\end{deluxetable}

Having identified some (limited) correlation between the simulated flux rope ejections and observed CME sources on timescales of a month or longer, we now consider the origin of this correlation.

The locations where flux ropes form in the simulation are determined by the structure of the magnetic field. Flux ropes form above polarity inversion lines where axial fields build up after flux cancellation \citep[see][]{yeates2009b}. The axial (sheared) components originate both from the emergence of twisted bipoles and from shearing by large-scale surface motions, so that the distribution of flux ropes in the simulations depends to some extent on the balance of these two contributions.

\begin{figure}[!htb]
\includegraphics[width=0.5\textwidth]{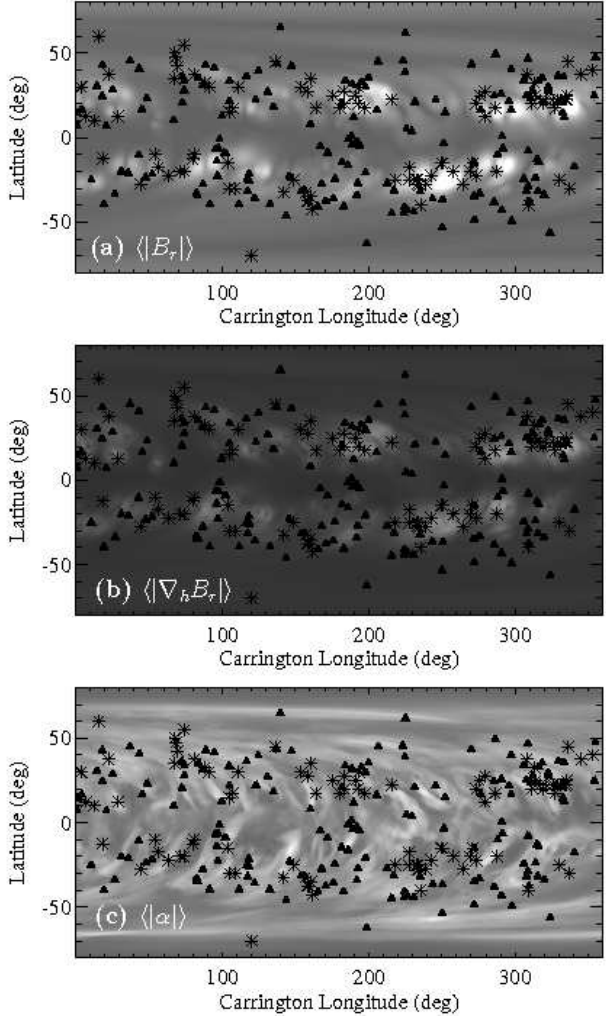}
\caption{Locations of simulated ejections and observed CME sources, overlaid on plots of magnetic field properties averaged over the comparison period: (a) radial surface field strength $\langle\left|B_r \right| \rangle$; (b) horizontal gradient in radial surface field $\langle\left|\nabla_hB_r\right| \rangle$; (c) current helicity $\langle|\alpha|\rangle$ at height $r=1. 03R_\odot$ in run A4. In each case asterisks show observed CME sources and triangles show locations of simulated ejections in run A4. }\label{fig:br}
\end{figure}

From the dashed line in Figure \ref{fig:time}, it is clear that the correlation disappears in simulation run AN, where no new bipolar regions were inserted after the initial condition. This suggests that the correlation observed in the other simulation runs arises primarily from the magnetic field distribution on the solar surface. This is essentially captured by the surface flux transport component of the simulations. To investigate this idea further, we consider purely the distribution of $B_r$ on the solar surface $r=R_\odot$, which is the same in all simulation runs except AN. This distribution represents the observational input driving the coronal magnetic field evolution in our model. It is summarized in Figure \ref{fig:br}(a), which shows the latitude-longitude distribution of $\langle|B_r|\rangle$, where the average is taken over all 136 days in the comparison period. The magnetic field is clearly non-uniform over the solar surface, and is concentrated in several major activity complexes, corresponding well to the clusters of observed CME sources identified in EUV (shown by asterisks). For comparison, the triangles show the locations of flux rope ejections in run A4, which are more evenly spread over the solar surface than the observed source regions. In a similar vein, Figure \ref{fig:br}(b) shows a map of $\langle|\nabla_hB_r|\rangle$, the time-averaged horizontal gradient of $B_r$ on the solar surface. The distribution resembles that of $\langle|B_r|\rangle$, except that there is a greater concentration in the active region belts relative to higher latitudes.

We may consider using these quantities $\langle|B_r|\rangle$ and $\langle|\nabla_hB_r|\rangle$ themselves as predictors of CME source locations. For this purpose we use the same binning as before, calculating the mean values of these new quantities in each bin. Table \ref{tab:br} shows the resulting correlations with observed CME source locations, for various choices of time bin size. Notably, both  $\langle|B_r|\rangle$ and $\langle|\nabla_hB_r|\rangle$ are better correlated with observed CME sources than are the flux rope ejections in any of the simulation runs. For example, the spatial distribution (136-day bins) of EUV sources has a correlation coefficient of $0.62$ with $\langle|B_r|\rangle$ and $0.67$ with $\langle|\nabla_hB_r|\rangle$, but only 0.49 with flux rope ejections in run A4. From Table \ref{tab:br}, we see that the correlations are maintained for time bins as short as 17 days, and are consistently better than the simulated ejections (Table \ref{tab:spatial}). Table \ref{tab:br} also shows that $\langle|\nabla_hB_r|\rangle$ is better correlated with observed CME locations than is $\langle|B_r|\rangle$. This is presumably because the former is more concentrated in active regions, where the majority of observed CME sources are located. Indeed, measures of active region complexity based on strong gradients of radial photospheric magnetic flux have been found to correlate with production from those regions both of large flares \citep{schrijver2009} and of CMEs \citep{falconer2008}. If vector magnetograms are available, alternative characterisations of the magnetic complexity are possible; for example, \citet{hahn2005} find a correlation between magnetic twist and the locations of H$\alpha$ flare signatures within active regions. Of course, although $\langle|\nabla_hB_r|\rangle$ is better correlated with the distribution of observed CMEs, it could not be used to predict their time of occurrence, because the photospheric magnetic field does not change significantly between the pre-eruption and post-eruption state. Therefore, time-dependent models of the 3D coronal magnetic field evolution must still be developed for this purpose.

\citet{yeates2009b} showed that the amount of helicity in emerging bipolar regions does alter the simulated flux rope ejections, but this would appear to have only a small effect on the correlation with observed CME sources. Figure \ref{fig:br}(c) shows the time-averaged current helicity $\alpha=\mathbf{j}\cdot\mathbf{B}/B^2$ in simulation run A4, at a height of $r=1.03R_\odot$ in the low corona \citep[see][for a discussion of the distribution of $\alpha$]{yeates2008b}. From the triangles in Figure \ref{fig:br}(c), we see that the ejections in run A4 always occur at locations of strong $\langle|\alpha|\rangle$ in the simulation. However, if we compute the correlation between the binned distribution of $\langle|\alpha|\rangle$ and the observed CME sources, we find a correlation coefficient of only $0.38$, much lower than for $\langle|B_r|\rangle$ or $\langle|\nabla_hB_r|\rangle$, and also lower than for the simulated ejections. This is because the observed CME sources are much more strongly concentrated in active regions than is the distribution of current helicity in the simulation. However, there are some EUV sources observed at higher latitudes, mainly quiescent filament eruptions. So the clustering of observed CME sources at active latitudes does not preclude the presence of coronal currents at higher latitudes. It is possible that there are two essentially different subsets of CMEs: those produced by quasi-static shearing and flux cancellation, not dissimilar to the distribution of flux rope ejections in our model, and a second set of events concentrated in active regions and resulting from more dynamic and/or smaller-scale processes not included in our model. Indeed, observations over several decades suggest that CMEs originating from active regions and associated with flares have different kinematic properties from those associated with prominence eruptions outside active regions. The former are faster, while the latter start more slowly and show gradual acceleration with height \citep{gosling1976,macqueen1983,sheeley1999}. However, it is still not certain whether CMEs with different speeds are qualitatively different \citep{gopalswamy2006}, and \citet{low2002} show how both could arise from flux ropes in different configurations. The EUV observations in this paper do not generally allow us to directly establish the presence or absence of a flux rope morphology, except for well-observed filament eruptions.

\section{Conclusion} \label{sec:conclusion}

In this paper, magnetic flux rope ejections in a global model of the coronal magnetic field evolution have been compared directly with low-coronal CME source regions identified in EUV, over a 136-day period in 1999. Despite limitations in identifying the source regions of observed CMEs, which was unambiguously possible for only a third of events, we find some definite correlation between the distributions of simulated ejections and observed CME sources. However, there are also clear differences between the observed and simulated distributions; the model produces fewer ejections overall, and much less clustering of ejections in active regions, even given the incomplete observations.

The correlation that is present between simulated and observed ejections originates primarily in the large-scale distribution of magnetic field on the solar surface, the basic observational input driving the coronal simulations. The correlation is only significant when the ejections are binned in time bins of at least 34 days, which is approximately the frequency (27 days) at which new observations of the surface magnetic field are are available as input to coronal simulations. There is a weak dependence of the correlation on the helicity in emerging active regions, but this is a secondary effect. Interestingly, the choice of emerging helicity that performs best is simulation run A4, which is also the run that most accurately reproduced the observed pattern of filament chirality \citep{yeates2008a}. However, a better predictor of CMEs than the locations of flux rope ejections is found to be simply the radial magnetic field on the solar surface, and in particular its horizontal gradient. This quantity is readily computed from surface flux transport simulations alone (or, periodically, from observed line-of-sight magnetic maps). Its good correlation with the observed CME source locations arises because the observed sources are clustered strongly into active regions, where the magnetic field is strong. By contrast, the simulated flux rope ejections are more evenly distributed over the solar surface, following the distribution of current helicity in the model. They do show some concentration towards active latitudes, reflecting the smaller scales of current helicity found in the more complex magnetic fields there, but the distribution is clearly different.

Our results allow us to place constraints on the possible initiation mechanisms of CMEs, at least in the rising phase of the 11-year Solar Cycle. Since flux ropes in the simulations are formed by gradual, quasi-static shearing of the magnetic field by large-scale motions, along with flux cancellation, we conclude that this mechanism cannot be responsible for all CMEs on the Sun, although it is sufficient to produce a significant fraction \citep{low2001,gibson2006}. Whether this fraction varies over the Solar Cycle is a question for future research. Based on our results here, we propose that there are two subsets of CMEs produced by essentially different mechanisms: those produced by large-scale, gradual transport of helicity (which may originate either from active regions or shearing by surface motions), and a second population concentrated in active regions. The former occur on timescales of weeks or even months, and are simulated in our model. The latter must occur on much shorter timescales, connected with the dynamic emergence of magnetic fields in active regions and their energetic restructuring. Since our model was originally developed to study the large-scale transport of magnetic helicity in the corona, it cannot produce the second type of CME using the present form of input data (synoptic magnetograms). To reproduce the frequent ejections observed from active regions will require, as a minimum, magnetogram data at much higher spatial and time resolution for input to the quasi-static simulations, at least within active regions themselves. However, given the dynamic nature of many events, it is likely that detailed, time-dependent MHD simulations of complex structures will need to be performed, based on observations of flux emergence such as those by {\it Hinode} \citep{lites2009}. At present, developing a detailed understanding of the initiation of such CMEs that could possibly lead to predictive capability is far out of reach, although there have been recent advances in modelling flux emergence in simplified configurations \citep{archontis2008}.

Perhaps a more achievable goal in the short term is to understand the evolution leading to the initiation of the first type of CME, those produced by quasi-static shearing. Above the active latitudes, remnant magnetic fields from active regions do not retain most of the complexity of their originating regions, and shearing motions along with flux cancellation over longer time-scales become more important. Our existing model provides a starting point to understand the net effect of these motions on flux rope development. However, the transport of helicity over the solar surface means that, at least at mid-latitudes, the formation of flux ropes is sensitive to the amount of magnetic helicity in individual emerging active regions, which is not yet routinely observed, despite recent improvements in techniques \citep{demoulin2009}. The forthcoming launch of NASA's Solar Dynamics Observatory (SDO) should improve the situation, with full-disk vector magnetograms at high cadence and resolution. This should allow us to put better constraints on the amount of helicity in the solar corona, and to further validate our quasi-static model for flux rope formation. There is also an open question as to the origin of sheared fields at high latitudes. In our present simulations, where helicity emerges from the solar interior only in active regions, differential rotation is the dominant source of helicity at high latitudes, but produces the opposite sign of helicity there to that observed \citep{yeates2008b}.

Finally, we note that the observational identification of CME source regions is not straightforward. Our attempt to use an alternative database of CME source regions in H$\alpha$ highlights the need for careful case-by-case investigation, preferably in multiple wavelengths. With a maximum cadence of 12 minutes in EIT 195\AA{} data, we were able to confidently determine the CME source regions of only 98 events on-disk, with a further 55 originating from behind the limb, out of a total of 330 CMEs. This is certainly incomplete. The situation should be improved in future, using higher-cadence observations at multiple EUV wavelengths, for example from SDO.  However, if future aims are to identify the relative importance of different CME initiation mechanisms in the global context, then a comprehensive catalog of CME sources must be built up over a long period. Only then can more definitive comparisons with theoretical models be made. Again, SDO offers some promise in this respect: for example, automated feature-finding algorithms for detecting the low-coronal signatures of CMEs in the anticipated SDO data are in preparation \citep{martens2009,attrill2009b}.

\acknowledgments
We thank M.J. Wills-Davey for useful suggestions. ARY was supported by NASA/LWS grant NNG05GK32G and contract NNM07AB07C to SAO. GDRA acknowledges the NASA/ROSES grant NNX09AB11G, and DN support from the Department of Science and Technology of the Government of India through the Ramanujan Fellowship. Simulations used the UKMHD parallel computer in St Andrews (funded jointly by SRIF/STFC), and support from the Royal Society through a research grant to DHM. We acknowledge the use of data from the LASCO and EIT consortia (SOHO is a project of international cooperation between ESA and NASA), and of the online EIT MPEG Movies Archive maintained by F. Auch\`{e}re. The CDAW CME catalog is generated and maintained at the CDAW Data Center by NASA and The Catholic University of America in cooperation with the Naval Research Laboratory. The Solar Geophysical Database is maintained by NOAA and made available courtesy of the Solar-Terrestrial Physics Division. Synoptic magnetogram data from NSO/Kitt Peak were produced cooperatively by NSF/NOAO, NASA/GSFC, and NOAA/SEL, and made available on the World Wide Web.

\appendix
\section{Observational Event List} \label{sec:app}

Table \ref{tab:data} gives the list of 330 LASCO CMEs, with associated EUV source information where a source was found. This includes source quality, approximate location, and notes on the type of signature(s) in each case.

\begin{deluxetable}{cccccccccl}
\setlength{\tabcolsep}{0.04in}
\tabletypesize{\scriptsize}
\tablecaption{Low coronal source regions for LASCO CMEs.}
\tablehead{\multicolumn{4}{c}{LASCO CME Data\tablenotemark{a}} & \multicolumn{4}{c}{Low Coronal EUV Source}\\
\colhead{Date} & \colhead{Time} & \colhead{CPA\tablenotemark{b}} & \colhead{Width\tablenotemark{c}} & \colhead{Quality\tablenotemark{d}} & \multicolumn{2}{c}{Longitude} & \multicolumn{2}{c}{Latitude} & \colhead{Notes\tablenotemark{e}} \\ 
&&&&& Min & Max & Min & Max &\\
\colhead{} & \colhead{(UT)} & \colhead{(deg)} & \colhead{(deg)} & \colhead{} & \colhead{(deg)} & \colhead{} & \colhead{(deg)} & \colhead{} & \colhead{} } 

\startdata
1999 May 13 & 23:26:07 & 311 & 105 & 2 & 50 & 70 & 15 & 25 & PEA near NW limb 21:35 \\
1999 May 14 & 03:06:05 & 55 & 63 & 2 & -90 & -70 & 40 & 50 & Activity \& dimming over NE limb 02:58 \\
1999 May 14 & 06:50:05 & 234 & 46 & 1 & 40 & 90 & -50 & -30 & Large QF eruption 03:22+ \\
1999 May 16 & 17:51:35 & 33 & 45 & 2 & 60 & 90 & -20 & 10 & Activity in large loop structures on limb,\\
&&&&&&&&& maybe partial F eruption \\
1999 May 17 & 00:50:07 & 293 & 113 & 3 & \multicolumn{4}{c}{far-side NW} & Eruption-opened loop structures above limb \\
1999 May 19 & 23:02:31 & 274 & 17 & 4 \\
1999 May 19 & 23:02:31 & 57 & 94 & 4 \\
1999 May 20 & 08:26:05 & 47 & 69 & 4 \\
1999 May 20 & 13:28:21 & 313 & 119 & 1 & -20 & 0 & 25 & 50 & Clear eruption, dimming, \& PEA loops from\\
&&&&&&&&&  centre NH, 08:48+ \\
1999 May 20 & 16:26:05 & 21 & 111 & 1 & -45 & 0 & 30 & 65 & Restructuring in NH AR after previous euption,\\
&&&&&&&&&  PEA from 18:35 \\
1999 May 20 & 21:26:08 & 115 & 91 & 4 \\
1999 May 21 & 10:50:05 & 341 & 130 & 4 \\
1999 May 21 & 17:50:06 & 20 & 218 & 3 & \multicolumn{4}{c}{far-side NE} & PEA above NE limb 20:11 (also clear in He304) \\
1999 May 23 & 07:40:05 & 288 & 67 & ? \\
1999 May 23 & 19:06:01 & 50 & 21 & ? \\
1999 May 24 & 10:30:05 & 101 & 28 & ? \\
1999 May 24 & 17:07:31 & 39 & 44 & 4 \\
1999 May 25 & 05:06:05 & 81 & 12 & 4 \\
1999 May 25 & 07:26:51 & 103 & 35 & 3 & \multicolumn{4}{c}{far-side SE} & PEA above limb 07:35+ \\
1999 May 25 & 10:50:05 & 268 & 178 & 1 & 70 & 90 & -50 & -10 & Clear eruption on SW limb 10:23, dimming,\\
&&&&&&&&&  coronal wave, flare \\
1999 May 25 & 16:26:05 & 26 & 133 & 4 \\
1999 May 25 & 23:26:05 & 5 & 138 & 4 \\
1999 May 26 & 04:26:05 & 44 & 71 & 2 & -45 & -35 & 20 & 30 & Surge-like small eruption + flare from NE sector 02:35 \\
1999 May 26 & 05:26:05 & 106 & 51 & 3 & \multicolumn{4}{c}{far-side SE} & Dimming \& PEA over SE limb 05:11+ \\
1999 May 26 & 08:06:05 & 321 & 101 & 1 & 35 & 90 & 40 & 70 & Slow QF eruption near NW limb starts 23:47 on\\
&&&&&&&&&  May 25, dynamic phase starts 07:24, dimming, PEA \\
1999 May 26 & 20:26:05 & 38 & 17 & 2 & -60 & -40 & 15 & 30 & Strong brightening in AR near NE limb 19:43+ \\
1999 May 26 & 22:26:05 & 110 & 70 & 1 & -90 & -35 & -60 & -20 & QF on SE limb starts to erupt 19:14, \\
&&&&&&&&& slow eruption until disappearance at 22:35 \\
\enddata

\tablenotetext{a}{As obtained from the CDAW catalogue at  \url{http://cdaw.gsfc.nasa.gov/CME\_list/}.}
\tablenotetext{b}{Central position angle.}
\tablenotetext{c}{Angular width in plane of sky.}
\tablenotetext{d}{Quality 1--4 of the EUV source identification, as defined in Section \ref{sec:obs}. The symbol ? indicates that EIT data were unavailable at that time.}
\tablenotetext{e}{Abbreviations used include AR (active region), PEA (post-eruption arcade), and QF (quiescent filament).}

\tablecomments{Table \ref{tab:data} is published in its entirety in the 
electronic edition of the {\it Astrophysical Journal} (or may obtained by email request to the first author).  A portion is 
shown here for guidance regarding its form and content.}

\label{tab:data}
\end{deluxetable}



\begin{thebibliography}{}

\bibitem[\protect\citeauthoryear{Amari et al.}{1996}]{amari1996}
Amari, T., Luciani, J.F., Aly, J.J, \& Tagger, M. 1996, ApJ, 306, 913

\bibitem[\protect\citeauthoryear{Amari et al.}{2003}]{amari2003}
Amari, T., Luciani, J.F., Aly, J.J, Mikic, Z., \& Linker, J. 2003, ApJ, 585, 1073

\bibitem[\protect\citeauthoryear{Antiochos et al.}{1999}]{antiochos1999}
Antiochos, S.K., DeVore, C.R., \& Klimchuk, J.A. 1999, ApJ, 510, 485

\bibitem[\protect\citeauthoryear{Archontis}{2008}]{archontis2008}
Archontis, V. 2008, J. Geophys. Res., 113, A03S04

\bibitem[\protect\citeauthoryear{Attrill et al.}{2009}]{attrill2009}
Attrill, G.D.R., Engell, A.J., Wills-Davey, M.J., Grigis, P., \& Testa, P. 2009, ApJ, 704, 1296

\bibitem[\protect\citeauthoryear{Attrill \& Wills-Davey}{2009}]{attrill2009b}
Attrill, G.D.R. \& Wills-Davey, M.J. 2009, Sol. Phys., in press

\bibitem[\protect\citeauthoryear{Barnes}{2007}]{barnes2007}
Barnes, G. 2007, ApJ, 670, L53

\bibitem[\protect\citeauthoryear{Bieber \& Rust}{1995}]{bieber1995}
Bieber, J.W. \& Rust, D.M. 1995, ApJ, 453, 911

\bibitem[\protect\citeauthoryear{Brueckner et al.}{1995}]{brueckner1995}
Brueckner, G.E., et al. 1995, Sol. Phys., 162, 357

\bibitem[\protect\citeauthoryear{Cohen et al.}{2009}]{cohen2009}
Cohen, O., Attrill, G.D.R., Manchester, W.B., \& Wills-Davey, M.J. 2009, ApJ, 705, 587

\bibitem[\protect\citeauthoryear{Cohen et al.}{2007}]{cohen2007}
Cohen, O., et al. 2007, ApJ, 654, L163

\bibitem[\protect\citeauthoryear{Cook et al.}{2009}]{cook2009}
Cook, G., Mackay, D.H., \& Nandy, D. 2009, ApJ, 704, 1021

\bibitem[\protect\citeauthoryear{Cremades \& Bothmer}{2004}]{cremades2004}
Cremades, H. \& Bothmer, V. 2004, A\&A, 422, 307

\bibitem[\protect\citeauthoryear{D\'{e}laboudini\`{e}re et al.}{1995}]{delaboudiniere1995}
D\'{e}laboudini\`{e}re, J.-P., et al. 1995, Sol. Phys., 162, 291

\bibitem[\protect\citeauthoryear{D\'{e}moulin \& Pariat}{2009}]{demoulin2009}
D\'{e}moulin, P. \& Pariat, E. 2009, Adv. Space Res., 43, 1013

\bibitem[\protect\citeauthoryear{Falconer et al.}{2008}]{falconer2008}
Falconer, D.A., Moore, R.L., \& Gary, G.A. 2008, ApJ, 689, 1433

\bibitem[\protect\citeauthoryear{Forbes}{2000}]{forbes2000}
Forbes, T.G. 2000, J. Geophys. Res., 105, A23153

\bibitem[\protect\citeauthoryear{Forbes \& Isenberg}{1991}]{forbes1991}
Forbes, T.G. \& Isenberg, P.A. 1991, ApJ, 373, 294

\bibitem[\protect\citeauthoryear{Gaizauskas et al.}{1983}]{gaizauskas1983}
Gaizauskas, V., Harvey, K.L., Harvey, J.W., \& Zwaan, C. 1983, ApJ, 265, 1056

\bibitem[\protect\citeauthoryear{Gibson \& Fan}{2006}]{gibson2006}
Gibson, S.E. \& Fan, Y. 2006, J. Geophys. Res., 111, A12103

\bibitem[\protect\citeauthoryear{Gopalswamy et al.}{2006}]{gopalswamy2006}
Gopalswamy, N., Miki\'{c}, Z., Maia, D., Alexander, D., Cremades, H., Kaufmann, P., Tripathi, D., \& Wang, Y.-M. 2006, Space Sci. Rev., 123, 303

\bibitem[\protect\citeauthoryear{Gopalswamy et al.}{2003}]{gopalswamy2003}
Gopalswamy, N., Shimojo, M., Lu, W., Yashiro, S., Shibasaki, K., \& Howard, R.A. 2003, ApJ, 586, 562

\bibitem[\protect\citeauthoryear{Gosling et al.}{1974}]{gosling1974}
Gosling, J.T., Hildner, E., MacQueen, R.M., Munro, R.H., Poland, A.I., \& Ross, C.L. 1974, J. Geophys. Res., 79, 4581

\bibitem[\protect\citeauthoryear{Gosling et al.}{1976}]{gosling1976}
Gosling, J.T., Hildner, E., MacQueen, R.M., Munro, R.H., Poland, A.I., \& Ross, C.L. 1976, Sol. Phys., 48, 389

\bibitem[\protect\citeauthoryear{Hahn et al.}{2005}]{hahn2005}
Hahn, M., Gaard, S., Jibben, P., Canfield, R.C., \& Nandy, D. 2005, ApJ, 629, 1135

\bibitem[\protect\citeauthoryear{Howard et al.}{2008}]{howard2008}
Howard, T.A., Nandy, D., \& Koepke, A.C. 2008, J. Geophys. Res., 113, A01104

\bibitem[\protect\citeauthoryear{Hudson et al.}{2006}]{hudson2006}
Hudson, H.S., Bougeret, J.-L., \& Burkepile, J. 2006, Space Sci. Rev., 123, 13

\bibitem[\protect\citeauthoryear{Isenberg et al.}{1993}]{isenberg1993}
Isenberg, P.A., Forbes, T.G., \& D\'{e}moulin, P. 1993, ApJ, 417, 368

\bibitem[\protect\citeauthoryear{Klimchuk}{2001}]{klimchuk2001}
Klimchuk, J.A. 2001, in Space Weather, ed. P. Song, H. Singer, \& G. Siscoe (Geophys. Monogr. 125; Washington: AGU), 143

\bibitem[\protect\citeauthoryear{Leka et al.}{1996}]{leka1996}
Leka, K.D., Canfield, R.C., McClymont, A.N., \& van Driel-Gesztelyi, L. 1996, ApJ, 462, 547

\bibitem[\protect\citeauthoryear{Li \& Luhmann}{2006}]{li2006}
Li, Y. \& Luhmann, J. 2006, ApJ, 648, 732

\bibitem[\protect\citeauthoryear{Lin \& van Ballegooijen}{2005}]{lin2005}
Lin, J. \& van Ballegooijen, A.A. 2005, ApJ, 629, 582

\bibitem[\protect\citeauthoryear{Lionello et al.}{2009}]{lionello2009}
Lionello, R., Linker, J.A., \& Miki\'{c}, Z. 2009, ApJ, 690, 902

\bibitem[\protect\citeauthoryear{Lites}{2009}]{lites2009}
Lites, B.W. 2009, Space Sci. Rev., 144, 197

\bibitem[\protect\citeauthoryear{Low}{2001}]{low2001}
Low, B.C. 2001, J. Geophys. Res., 106, A25141

\bibitem[\protect\citeauthoryear{Low \& Zhang}{2002}]{low2002}
Low, B.C. \& Zhang, M. 2002, ApJ, 564, L53

\bibitem[\protect\citeauthoryear{Mackay \& van Ballegooijen}{2005}]{mackay2005}
Mackay, D.H. \& van Ballegooijen, A.A. 2005, ApJ, 621, L77

\bibitem[\protect\citeauthoryear{Mackay \& van Ballegooijen}{2006}]{mackay2006a}
Mackay, D.H. \& van Ballegooijen, A.A. 2006, ApJ, 641, 577

\bibitem[\protect\citeauthoryear{MacQueen \& Fisher}{1983}]{macqueen1983}
MacQueen, R.M. \& Fisher, R.R. 1983, Sol. Phys., 89, 89

\bibitem[\protect\citeauthoryear{Manchester et al.}{2008}]{manchester2008}
Manchester, W.B., et al. 2008, ApJ, 684, 1448

\bibitem[\protect\citeauthoryear{Martens et al.}{2009}]{martens2009}
Martens, P.C.H., et al. 2009, Sol. Phys., submitted.

\bibitem[\protect\citeauthoryear{Miki{\'c} \& Linker}{1994}]{mikic1994}
Miki{\'c}, Z. \& Linker, J.A. 1994, ApJ, 430, 898

\bibitem[\protect\citeauthoryear{Nandy et al.}{2008}]{nandy2008}
Nandy, D., Mackay, D.H., Canfield, R.C., \& Martens, P.C.H. 2008, J. Atmos. Sol.-Terr. Phys., 70, 605 

\bibitem[\protect\citeauthoryear{Owens et al.}{2007}]{owens2007}
Owens, M.J., Schwadron, N.A., Crooker, N.U., Hughes, W.J., \& Spence, H.E. 2007, Geophys. Res. Lett., 34, L06104

\bibitem[\protect\citeauthoryear{Pevtsov et al.}{1995}]{pevtsov1995}
Pevtsov, A.A., Canfield, R.C., \& Metcalf, T.R. 1995, ApJ, 440, L109

\bibitem[\protect\citeauthoryear{Plunkett et al.}{2001}]{plunkett2001}
Plunkett, S.P., Thompson, B.J., St. Cyr, O.C., \& Howard, R.A. 2001, J. Atmos. Sol.-Terr. Phys., 63, 389 

\bibitem[\protect\citeauthoryear{Pneuman}{1983}]{pneuman1983}
Pneuman, G.W. 1983, Sol. Phys., 88, 219

\bibitem[\protect\citeauthoryear{Pojoga \& Huang}{2003}]{pojoga2003}
Pojoga S. \& Huang, T.S. 2003, Adv. Space Res., 32, 2641

\bibitem[\protect\citeauthoryear{Riley et al.}{2006}]{riley2006}
Riley, P., Linker, J.A., {Miki{\'c}}, Z., Lionello, R., Ledvina, S.A., \& Luhmann
, J.G. 2006, ApJ, 653, 1510

\bibitem[\protect\citeauthoryear{Riley et al.}{2008}]{riley2008}
Riley, P., Lionello, R., Miki\'{c}, Z., \& Linker, J.A. 2008, ApJ, 672, 1221

\bibitem[\protect\citeauthoryear{Robbrecht \& Berghmans}{2004}]{robbrecht2004}
Robbrecht, E. \& Berghmans, D. 2004, A\&A, 425, 1097

\bibitem[\protect\citeauthoryear{Robbrecht et al.}{2009a}]{robbrecht2009a}
Robbrecht, E., Berghmans, D., \& van der Linden, R.A.M. 2009a, ApJ, 691, 1222

\bibitem[\protect\citeauthoryear{Robbrecht et al.}{2009b}]{robbrecht2009b}
Robbrecht, E., Patsourakos, S., \& Vourlidas, A. 2009b, ApJ, 701, 283

\bibitem[\protect\citeauthoryear{Schrijver}{2009}]{schrijver2009}
Schrijver, C.J. 2009, Adv. Space Res., 43, 739

\bibitem[\protect\citeauthoryear{Schrijver et al.}{2008}]{schrijver2008}
Schrijver, C.J., et al. 2008, ApJ, 675, 1637

\bibitem[\protect\citeauthoryear{Schwenn}{2006}]{schwenn2006}
Schwenn, R. 2006, Living Rev. Sol. Phys., 3, 2 (\url{http://www.livingreviews.org/lrsp-2006-2})

\bibitem[\protect\citeauthoryear{Sheeley}{2005}]{sheeley2005}
Sheeley, N.R. 2005, Living Rev. Sol. Phys., 2, 5 (\url{http://www.livingreviews.org/lrsp-2005-5})

\bibitem[\protect\citeauthoryear{Sheeley et al.}{1999}]{sheeley1999}
Sheeley, N.R., Walters, J.H., Wang, Y.-M., \& Howard, R.A. 1999, J. Geophys. Res., 104, 24739

\bibitem[\protect\citeauthoryear{Ugarte-Urra et al.}{2007}]{ugarteurra2007}
Ugarte-Urra, I., Warren, H.P., \& Winebarger, A.R. 2007, ApJ, 662, 1293

\bibitem[\protect\citeauthoryear{van Ballegooijen et al.}{1998}]{vanballegooijen1998}
van Ballegooijen, A.A., Cartledge, N.P., \& Priest, E.R. 1998, ApJ, 501, 866

\bibitem[\protect\citeauthoryear{van Ballegooijen \& Martens}{1989}]{vanballegooijen1989}
van Ballegooijen, A.A. \& Martens, P.C.H. 1989, ApJ, 343, 971

\bibitem[\protect\citeauthoryear{van Ballegooijen \& Martens}{1990}]{vanballegooijen1990}
van Ballegooijen, A.A. \& Martens, P.C.H. 1990, ApJ, 361, 283

\bibitem[\protect\citeauthoryear{van Ballegooijen et al.}{2000}]{vanballegooijen2000}
van Ballegooijen, A.A., Priest, E.R., \& Mackay, D.H. 2000, ApJ, 539, 983

\bibitem[\protect\citeauthoryear{Yang et al.}{1986}]{yang1986}
Yang, W.H., Sturrock, P.A., \& Antiochos, S.K. 1986, ApJ, 309, 383

\bibitem[\protect\citeauthoryear{Yashiro et al.}{2004}]{yashiro2004}
Yashiro, S., Gopalswamy, N., Michalek, G., St. Cyr, O.C., Plunkett, S.P., Rich, N.B., \& Howard, R.A. 2004, J. Geophys. Res., 109, A07105 

\bibitem[\protect\citeauthoryear{Yashiro et al.}{2008}]{yashiro2008}
Yashiro, S., Michalek, G., \& Gopalswamy, N. 2008, Ann. Geophys., 26, 3103 

\bibitem[\protect\citeauthoryear{Yeates \& Mackay}{2009a}]{yeates2009a}
Yeates, A.R. \& Mackay, D.H. 2009a, Sol. Phys., 254, 77

\bibitem[\protect\citeauthoryear{Yeates \& Mackay}{2009b}]{yeates2009b}
Yeates, A.R. \& Mackay, D.H. 2009b, ApJ, 699, 1024

\bibitem[\protect\citeauthoryear{Yeates et al.}{2007}]{yeates2007}
Yeates, A.R., Mackay, D.H., \& van Ballegooijen, A.A. 2007, Sol. Phys., 245, 87

\bibitem[\protect\citeauthoryear{Yeates et al.}{2008a}]{yeates2008a}
Yeates, A.R., Mackay, D.H., \& van Ballegooijen, A.A. 2008a, Sol. Phys., 247, 103

\bibitem[\protect\citeauthoryear{Yeates et al.}{2008b}]{yeates2008b}
Yeates, A.R., Mackay, D.H., \& van Ballegooijen, A.A. 2008b, ApJ, 680, L165

\bibitem[\protect\citeauthoryear{Zhang \& Low}{2001}]{zhang2001}
Zhang, M. \& Low, B.C. 2001, ApJ, 561, 406

\bibitem[\protect\citeauthoryear{Zirker et al.}{1997}]{zirker1997}
Zirker, J.B., Martin, S.F., Harvey, K., \& Gaizauskas, V. 1997, Sol. Phys., 175, 27

\end{thebibliography}
\end{document}